\documentclass[fleqn,10pt]{SelfArx} % Document font size and equations flushed left

\definecolor{color1}{RGB}{0,0,90} % Color of the article title and sections
\definecolor{color2}{RGB}{0,20,20} % Color of the boxes behind the abstract and headings

\usepackage{hyperref} % Required for hyperlinks
\hypersetup{hidelinks,colorlinks,breaklinks=true,urlcolor=color2,citecolor=color1,linkcolor=color1,bookmarksopen=false,pdftitle={Title},pdfauthor={Author},urlcolor=blue}

\usepackage{graphicx}
\usepackage[square]{natbib}
\usepackage{amsmath}
\usepackage{multirow}
\usepackage{subfigure}

\newcommand{\n}[1]{\mathrm{#1}}

\JournalInfo{Published in International Journal of Thermal Sciences, Vol. 85, 12–20, 2014} % Journal information
\Archive{\href{http://dx.doi.org/10.1016/j.ijthermalsci.2014.06.003}{DOI: 10.1016/j.ijthermalsci.2014.06.003}} % Additional notes (e.g. copyright, DOI, review/research article)

\PaperTitle{Analysis of the internal heat losses in a thermoelectric generator} % Article title

\Authors{R. Bj\o{}rk, D. V. Christensen, D. Eriksen and N. Pryds} % Authors
\affiliation{\textit{Department of Energy Conversion and Storage, Technical University of Denmark - DTU, Frederiksborgvej 399, DK-4000 Roskilde, Denmark}} % Author affiliation
\affiliation{*\textbf{Corresponding author}: rabj@dtu.dk} % Corresponding author

 % Keywords - if you don't want any simply remove all the text between the curly brackets
\newcommand{\keywordname}{Keywords} % Defines the keywords heading name

\Abstract{A 3D thermoelectric numerical model is used to investigate different internal heat loss mechanisms for a thermoelectric generator with bismuth telluride p- and n-legs. The model considers all thermoelectric effects, temperature dependent material parameters and simultaneous convective, conductive and radiative heat losses, including surface to surface radiation. For radiative heat losses it is shown that for the temperatures considered here, surface to ambient radiation is a good approximation of the heat loss. For conductive heat transfer the module efficiency is shown to be comparable to the case of radiative losses. Finally, heat losses due to internal natural convection in the module is shown to be negligible for the millimetre sized modules considered here. The combined case of radiative and conductive heat transfer resulted in the lowest efficiency. The optimized load resistance is found to decrease for increased heat loss. The leg dimensions are varied for all heat losses cases and it is shown that the ideal way to construct a TEG module with minimal heat losses and maximum efficiency is to either use a good insulating material between the legs or evacuate the module completely, and use small and wide legs closely spaced.}

\begin{document}

\flushbottom % Makes all text pages the same height

\maketitle % Print the title and abstract box

%\tableofcontents % Print the contents section

\thispagestyle{empty} % Removes page numbering from the first page

\section{Introduction}
Power generation through the thermoelectric effect is a subject of increasing interest both scientifically and commercially. The performance of a thermoelectric generator (TEG) is often estimated using numerical modelling before actual physical modules are constructed. This allows for the development of new and better thermoelectric generators in a cost effective manner, as well as for designing and optimizing modules to achieve as high an efficiency as possible. To accurately design the most efficient module possible, a large number of factors must be taken into account including geometric parameters, temperature dependence of the material properties, heat losses and thermal and electrical contact resistances.

A substantial number of numerical models of thermoelectric generators, both of unicouples and modules, have been published, but most of these only consider the ideal case with no heat losses and contact resistances. While the latter effect presents some challenges to determine experimentally, it can easily be included in a numerical model, if the contact resistance is known from experimental measurements. Including heat losses in a numerical model is more troublesome, but is never the less important as heat losses can significantly degrade the performance of a TEG. Experimental thermoelectric generator output is known to deviate from a TEG model with no heat losses; for example have heat losses determined experimentally previously been reported to reduce the efficiency by about 40-50\% \citep{ElGenk_2006,Hung_2014}.

Heat losses have been considered in numerical models of TEGs in both one, two and three dimensional models. \citet{Muto_2009} considered a 1D TEG model where the radiative heat loss is given by Stefan-Boltzmann law for a specified perimeter length, whereas \citet{Meng_2011} consider a 1D model including all heat losses (radiative, conductive and convective) by assuming the temperature in the air gap between the TEG legs to be equal to that in the legs themselves. For three dimensions a substantial number of publications exists where a TEG is modelled, however without heat losses, using a finite element model (FEM) \citep{Kim_2009,Ebling_2010,Jang_2011,Seetawan_2012,Wang_2013a,Huang_2013,Wang_2014}. When heat losses are considered, most frequently the models reported in literature only consider a single heat loss mechanism and typically the heat loss mechanisms are simplified substantially. For general heat loss and for fluid flow problems often only heat loss through a simple heat transfer coefficient has been considered \citep{Harris_2006, Chen_2011, Reddy_2013, Bauknecht_2013, Yang_2011, Wang_2012a}. For heat loss through radiation, only surface to ambient radiation and not surface to surface radiation has been considered \citep{Saber_2002, Ziolkowski_2010, Yang_2011, Reddy_2013}. Using the simplified physical assumptions presented above, different heat loss mechanisms have been compared, where it was found that the convectional losses are greatest followed by radiation and thermal conduction of a solid filling material within the voids of the module \citep{Ziolkowski_2010}.

Here we consider the performance of a TEG, accounting for heat losses in full detail numerically by modelling surface to surface radiation as well as conductive and convective heat losses, where the flow field is completely modelled for the latter. For each heat loss mechanism the influence of both geometrical factors and physical parameters are investigated, with the aim of determining the influence of the various heat loss mechanisms on the performance of a TEG. This knowledge will allow for the optimal TEG with regards to cost and efficiency to be designed.

\section{Three dimensional TEG model}
A 3D thermoelectric model of a module with p- and n-type legs, electrodes and substrates has been set up and implemented in the commercial finite element software Comsol Multiphysics \citep{comsol}. The coupled differential equations describe the electrical current density, $\mathbf{J}$, and the heat flux, $\mathbf{J}_{Q}$, as \citep{Yang_2012}
\begin{eqnarray}\label{Eq.def}
-\mathbf{J} &=&\sigma\nabla{}V + \sigma\alpha\nabla T\nonumber\\
\mathbf{J}_{Q} &=& -\kappa\nabla T + T\alpha\mathbf{J}
\end{eqnarray}
Here $\sigma$ is the electrical conductivity, $V$ is the electrical potential, $\alpha$ is the Seebeck coefficient, $T$ is the temperature,  and $\kappa$ is the thermal conductivity. In the equation for $\mathbf{J}$, the first term is Ohm’s law, while the second describes the Seebeck effect. In the equation for $\mathbf{J}_{Q}$ the first term describes Fourier heat conduction, while the latter describes the Peltier effect. Numerically, the latter term in the equation for $\mathbf{J}$ is implemented as an ``External Current Density'' in Comsol, while the expression for $\mathbf{J}_{Q}$ is modified through a ``Weak Expression''\footnote{The latest version of Comsol, 4.4, has built-in support of thermoelectric materials. The implementation discussed here is equivalent to this implementation.}.

In steady state operation the current density is divergence-free
\begin{eqnarray}
\nabla{}\cdot{}\mathbf{J} &=&0~,
\end{eqnarray}
whereas the heat flux is given by
\begin{eqnarray}
\nabla{}\cdot{}\mathbf{J}_{Q} &=& -\nabla{}\cdot{}(V\mathbf{J})~.
\end{eqnarray}

For the case of steady state operation, the energy accumulation, $\dot{e}$, must be zero, i.e.
\begin{eqnarray}
\dot{e} &=& \nabla{}\cdot{}(\kappa{}\nabla{}T)-\nabla{}\cdot{}((V+T\alpha{})\mathbf{J}) = 0.
\end{eqnarray}
Remembering that the current is divergence-free the energy accumulation becomes
\begin{eqnarray}
\dot{e} &=& \nabla{}\cdot{}(\kappa{}\nabla{}T)-\mathbf{J}\cdot{}(\sigma^{-1}\mathbf{J})-T\nabla{}\alpha{}\cdot{}\mathbf{J}.
\end{eqnarray}
where the middle term is the Joule heating and the latter term is the Thomson effect, both of which are present in the simulations.

These equations are solved on a finite element mesh for the desired geometry and thermal and electrical boundary conditions. The geometry considered here is that of several unicouples connected electrically in serial and thermally in parallel into a module. Either the material properties of the electrodes must be specified or an infinitely good electrical contact can be assumed. An external electrical load resistance, $R_\n{ext}$, is applied to the TEG. This is modelled as a boundary condition, described as a thin sheet of resistive material, with conductivity $\sigma_\n{s}$ and thickness $d_\n{s}$, connected to a reference potential $V_\n{ref}=0$ in one end of the module. The equation for this boundary condition is
\begin{eqnarray}
\mathbf{n}\cdot{}(\mathbf{J}_1-\mathbf{J}_2) &=& \frac{\sigma_\n{s}}{d_\n{s}}(V-V_\n{ref})~,
\end{eqnarray}
where $\mathbf{n}$ is the normal vector to the surface. The other end of the module is assumed connected to ground, $V=0$. All surfaces of the TE legs not exposed to the electrodes are assumed to be electrically insulating, i.e. $\mathbf{J}\cdot{}\mathbf{n}=0$, while the current is conserved over all internal boundaries. Finally, a contact resistance can be specified on all boundaries, in order to model an imperfect joining between the legs and the electrodes. However, here we assume the joining to be perfect in all cases.

Thermally, any two parameters out of the hot side temperature, cold side temperature, hot side heat flux and cold side heat flux must be specified as input parameters to the model. The thermal boundary conditions depend on the heat losses modelled, e.g. whether an infinite module or a module of finite size, with heat loss through the sides of the module, is considered. Here, we consider modules where no heat is lost through the sides of the module, but only lost internally in the module.

\subsection{Radiative heat losses}
If radiative heat losses are considered, for a surface radiating through a transparent medium, the radiative heat flux is given by the difference between the incoming radiation and the radiation leaving the surface, i.e.
\begin{eqnarray}
Q=\epsilon{}(G-\sigma{}_\n{SB}T^4)
\end{eqnarray}
where $\epsilon$ is the emissivity, $G$ is the incoming radiative heat flux, or irradiation, and $\sigma{}_\n{SB}=5.67*10^{-8}$ Wm$^{-2}$K$^{-4}$ is Stefan-Boltzmanns constant. The irradiation can in general be written as
\begin{eqnarray}
G = G_\n{m} + F_\n{amb}\sigma{}_\n{SB}T_\n{amb}^4
\end{eqnarray}
where $G_\n{m}$ is the mutual irradiation from other surfaces, $F_\n{amb}$ is the ambient view factor, i.e. the fraction of the field of view that is not covered by other surfaces, and $T_\n{amb}$ is the ambient temperature. The mutual irradiation will depend on the total outgoing radiative flux, or radiosity, $J$, at every other point in view. This results in an implicit equation for the radiosity as
\begin{eqnarray}
J = (1-\epsilon)(G_\n{m}(J)+F_\n{amb}\sigma{}_\n{SB}T_\n{amb}^4)+\epsilon{}\sigma{}_\n{SB}T^4
\end{eqnarray}
where $G_\n{m}(J)$ is the mutual irradiation from other surfaces, which depends on $J$. For the case of blackbody radiation, the radiosity only depends on the temperature of the surface, as all incoming radiation is absorbed and converted to heat. These physics are built into and verified in the numerical software package Comsol used here \cite{comsol}. The heat transfer equation and the equation for the radiosity are solved in parallel. When finding the view factor, shadow effects are taken into account. The view factor is evaluated using a $z$-buffered projection on the side of a hemicube to account for shadowing effects. This is somewhat analogue to rendering digital images of the geometry in five different directions and counting the pixels in each mesh element to evaluate its view factor \cite{comsol}.

For radiative heat transfer the side walls of the module are assumed to be so-called diffuse mirrors. This is a common approximation of a surface that is well insulated on one side and for which convection effects can be neglected on the opposite (radiating) side. It resembles a mirror that absorbs all irradiation and then radiates it back in all directions in all points. The radiative heat flux from a diffuse mirror boundary is zero \cite{comsol}.

\subsection{Conductive and convective heat losses}
For conductive heat losses the heat loss equation solved is simply that of pure conductive heat transfer (Eq. (\ref{Eq.def}) for $\mathbf{J}_{Q}$ with $\mathbf{J}=0$). For convective heat losses the full compressible Navier-Stokes equation and the heat transfer equation are solved, i.e.
\begin{eqnarray}
\nabla{}\cdot{}(\rho{}\mathbf{u}) &=& 0\nonumber\\
\rho\mathbf{u}\cdot{}\nabla{}\mathbf{u} &=& -\nabla{}p+\nabla{}\cdot{}\left(\mu{}\left(\nabla{}\mathbf{u}+(\nabla{}\mathbf{u})^{T}\right)\times{}\right.\\
&&\left.-\frac{2}{3}\mu{}(\nabla{}\cdot{}\mathbf{u})\mathbf{I}\right)+\mathbf{F}
\end{eqnarray}
where $\rho{}$ is the density, $\mathbf{u}$ is the velocity vector, $p$ is the pressure, $\mu$ is the dynamic viscosity and the body force is given by $\mathbf{F}=\rho{}\mathbf{g}$. In this equation the $T$ denotes the transpose. In practice the pressure difference inside the TEG module will be small, which does allow the incompressible Navier-Stokes equation to be used, as the convection is only caused by a change in density with respect to temperature and not with pressure.

\subsection{Modelled system}
We consider a thermoelectric system consisting of a p-type nanocrystalline bulk Bi$_2$SbTe$_3$ \citep{Ma_2008} and n-type Bi$_2$Te$_3$ heat-treated nanocompound \citep{Kim_2012}, with temperature dependent experimentally measured properties. The materials represent some of the best performing bismuth telluride materials published to date. This choice of materials allows the computed results to be applicable for almost all commercial modules to date, as these consist of p- and n-type BiTe legs. In all simulations the hot side temperature is taken to be 523.15 K and the cold side to be 293.15 K, as these represent realistic operating temperatures for the BiTe materials, if any degradation performance of the BiTe over time at the high temperature is disregarded. A physical top and bottom plate is present in all simulations, with properties as given in Table \ref{Table.1}. In all models no contact resistance, neither electrically or thermally, is included. Also, for all models the electrical contacts between legs are assumed to be perfect. This means that the potential generated by one leg is directly applied to the next leg in the module. This removes the need to physically resolve and mesh the electrical contacts between the legs.

\section{Model verification}
Before considering heat losses, we first consider a system with no heat losses, in order to verify the model and to find the optimal geometry of the system used for analysing heat losses. We consider a system with a leg length of 1 mm and hot and cold side temperatures as given in Table \ref{Table.1}. For a TEG with no thermal losses, the efficiency, $\eta$, is defined as
\begin{eqnarray}\label{Eq.Efficiency_def}
\eta{}=\frac{P}{Q_\n{in}}
\end{eqnarray}
where $P$ is the power produced, and $Q_\n{in}$ is the heat flowing into the TEG, i.e. the sum of the heat flowing into the p- and n-leg, respectively. The power is simply given by $P=IV$, where $I$ is the current produced and $V$ is the total electric potential over the entire TEG module. The power per unit area is defined as the power produced divided with the total cross sectional area of the legs, i.e. $A_\n{n}+A_\n{p}$.

\begin{figure}[!t]
  \centering
  \includegraphics[width=1\columnwidth]{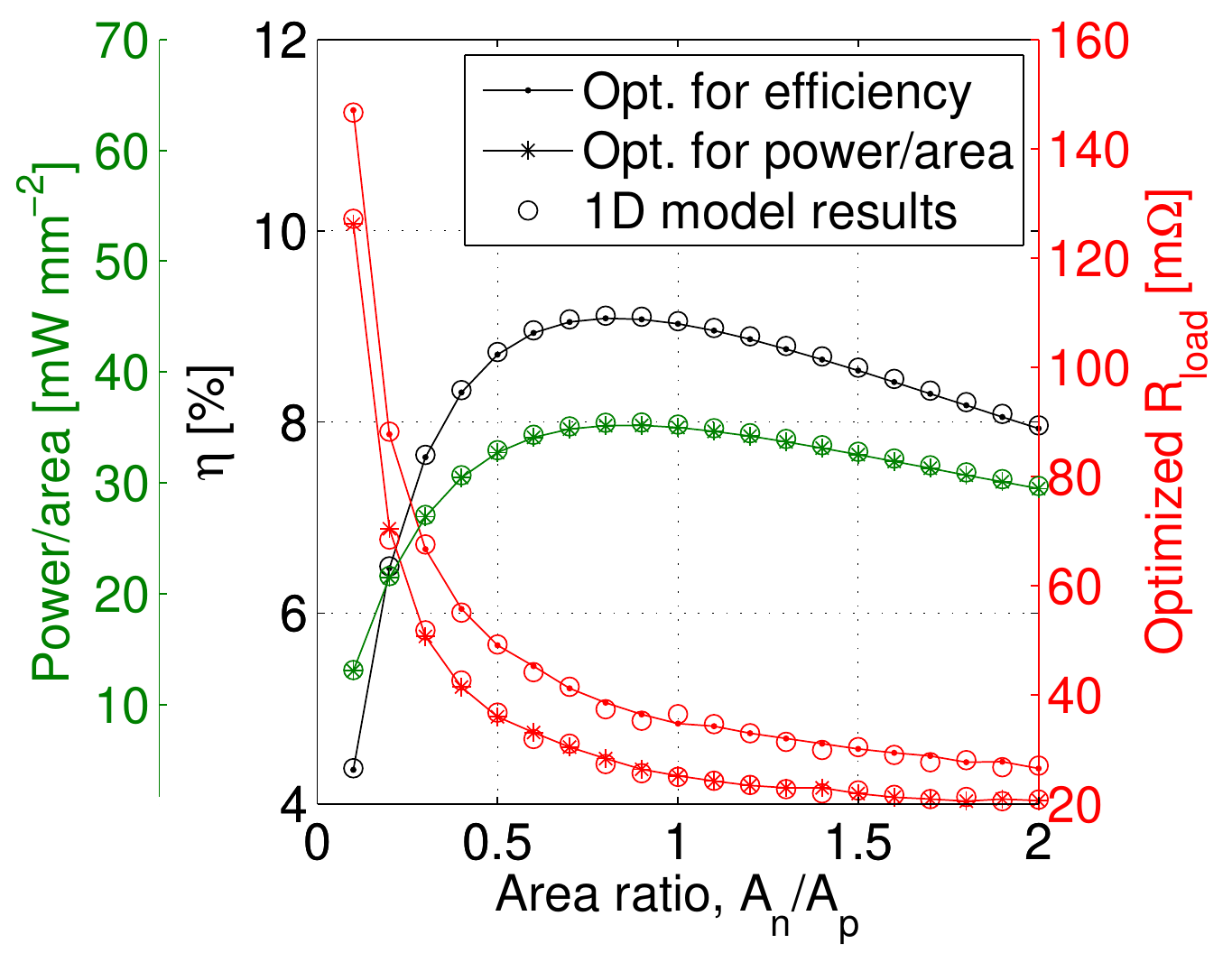}
  \caption{The optimized efficiency, $\eta$, and power per unit area and corresponding load resistances as function of area ratio. Also shown are results calculated using a 1D numerical model \citep{Snyder_2003,TE_Handbook_Ch9}.}
  \label{Fig_Line_no_losses_opt}
\end{figure}

For a system with a fixed temperature span, but without heat losses, the efficiency only depends on the area ratio between cross-sectional areas of the p- and n-legs, $A_\n{n}/A_\n{p}$, and the load resistance, $R_\n{ext}$. The efficiency and power per area as function of the area ratio are shown in Fig. \ref{Fig_Line_no_losses_opt} for the load resistance that results in the highest values. The optimum efficiency and power per area occur at an optimal area ratio of around $A_\n{n}/A_\n{p} =0.75$ in both cases. However, there is little decrease in the performance when slightly increasing the area ratio. In the following heat loss analysis we will fix the area ratio to $A_\n{n}/A_\n{p} = 1$, at which the no-heat-loss efficiency is 9\%, as this is the area ratio usually found in commercial modules.

As can be seen from Fig. \ref{Fig_Line_no_losses_opt} there is an excellent agreement between the model presented here and a 1D model of a TEG \citep{Snyder_2003,TE_Handbook_Ch9} for the case of no heat losses \citep{Fraisse_2013}. The 1D model reformulates the general equation for the efficiency of a TEG, allowing for simple integration to determine the thermoelectric state variables through a unicouple. For the area ratio of $A_\n{n}/A_\n{p} = 1$, a detailed investigation of the temperature profile in both legs reveals this to be  close to linear, with variation less than 3.5 K when comparing the same position in the n- and p-leg. This means that ideally, when regarding heat losses, placing the legs very closely together does not result in a large leg-to-leg heat transfer.

\section{Heat losses mechanisms}
We will consider heat loss by radiation, conduction and convection, both singlehandedly and simultaneously, as these are the heat loss mechanisms for a TEG. As previously mentioned we consider legs with dimensions of 1$\times$1$\times$1 mm$^3$, which is close to the optimal area ratio and to the dimensions used in commercial TEG modules \citep{Marlow_2013}. We have not optimized the cross-sectional area ratio between n- and p-legs for the different heat loss cases, as we wish to compare the magnitude of the different heat loss mechanisms for identical TEGs. For all modelled systems the external load resistance that optimizes the efficiency for the given system is found.

\begin{figure*}[p]
  \centering
  \includegraphics[width=1.8\columnwidth]{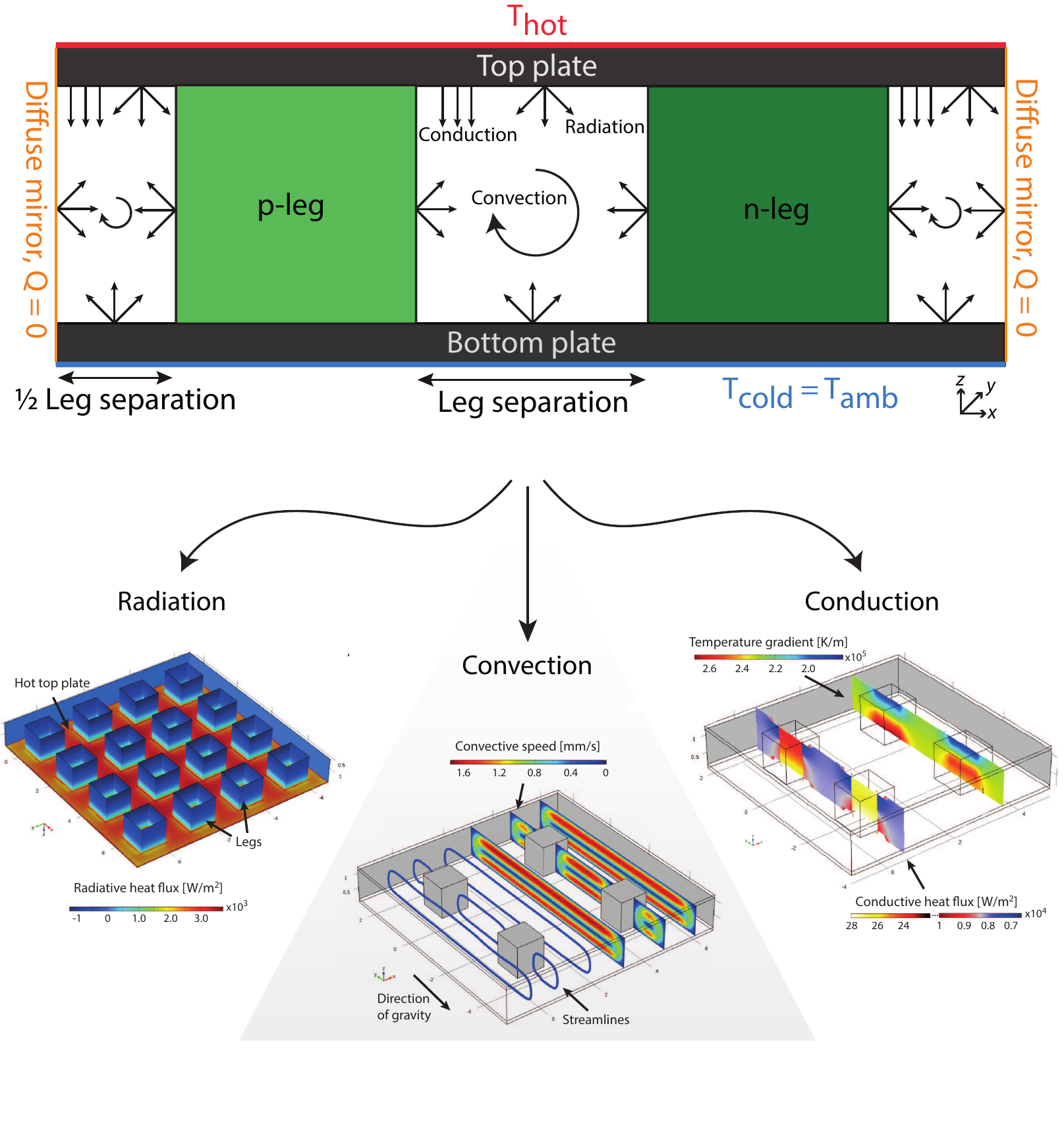}
  \caption{An illustration of a slice of the thermoelectric system modelled for the different heat loss mechanisms considered. As previously mentioned the electrical connection between the legs is assumed to be perfect, removing the need to resolve the electrical contacts. The full 3D system is modelled. The distance between the legs is the same in the $x$ and $y$ directions. The different heat loss mechanisms modelled are also illustrated in the figure. The radiative heat flux for a 4x4 module including surface to surface radiation is shown for a leg separation distance of 1 mm, with the TEG viewed from the bottom. For the case of convection the flow field is shown for a 3 mm leg separation 2x2 module filled with air and including surface to surface radiation, conduction and convection. Finally the conductive heat flux and the temperature gradient is shown for the case of a 2 mm leg separation 2x2 module with a solid thermal insulator placed between the legs.}
  \label{Fig_losses_ill_combined_single}
\end{figure*}

We consider whole modules, consisting of $n\times{}n$-legs, equidistantly spaced and topped with a hot plate and with a cold bottom plate. In some cases the efficiency is the same regardless of the value of $n$, but if this is not the case we consider modules with $2\times{}2$ or $4\times{}4$ legs. Such a case could be surface to surface radiation, where the number of visible legs changes depending on the total number of legs. The legs are separated by a distance, termed the leg separation distance, and the top and bottom plates extend half of this distance out from the legs, as also shown in Fig. \ref{Fig_losses_ill_combined_single}. In this way the area of the legs compared to the area of the plates does not vary based on the number of legs in the module.

The module has adiabatic side walls. Thus there is no external heat loss from the module, but only internal losses, e.g. heat passing directly from the hot plate to the cold plate, bypassing the legs. Modelling heat loss through the side walls would require knowing the conditions outside of the TEG, i.e. simulating the precise environment that the TEG is placed in, which would mean that the heat losses could not be generally calculated. An illustration of a slice through the 3D system modelled is shown in Fig. \ref{Fig_losses_ill_combined_single}, as well as some illustrations of the different heat transfer mechanisms considered.

\begin{table*}
\begin{tabular}{c | c c c}
Module & $T_\n{hot} = 523.15$ K & $T_\n{cold} = 293.15$ K & $\epsilon=1$ \\ \hline
Legs and plates & $x_\n{legs} = y_\n{legs} = z_\n{legs}$ = 1 mm & $z_\n{plates}$ = 0.1 mm & $\kappa{}_\n{plates}=429$ Wm$^{-1}$K$^{-1}$
\end{tabular}
\caption{The general module parameters. The thermal conductivity of the plates is identical to Ag at room temperature.} \label{Table.1}
\end{table*}

For a module with heat losses, we will consider the system efficiency in the same way as for the no heat loss case, i.e. using Eq. (\ref{Eq.Efficiency_def}), but where $Q_\n{in}$ is taken as the total heat that is flowing from the hot side. This means that for e.g. a system in which the legs are embedded in a thermal insulator, the heat flowing directly from the hot side to the cold side through the thermal insulator, bypassing the legs completely, is included in the efficiency. Thus the efficiency measures the fraction of heat deducted from the hot side that is transformed into electricity by the TEG.

\begin{figure}[!t]
  \centering
  \includegraphics[width=1\columnwidth]{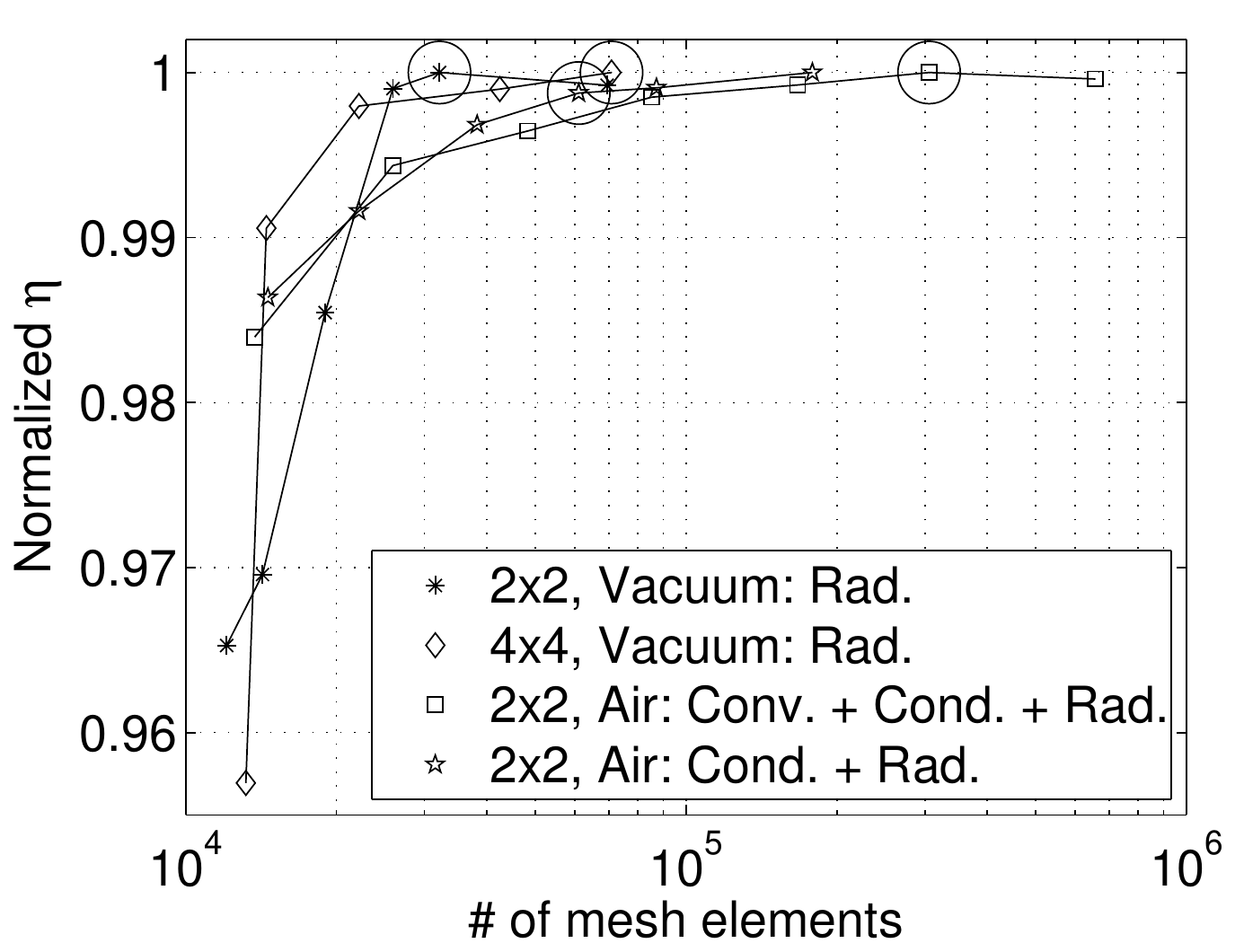}
  \caption{The normalized efficiency as function of the number of mesh elements of the finite element mesh for some of the different heat loss mechanisms and different number of legs considered. The chosen mesh resolutions have been indicated for the different cases with a circle.}
  \label{Fig_Mesh_analysis}
\end{figure}

For each model containing different heat loss mechanisms a mesh size analysis was conducted to ensure that the results were not a function of mesh size. The normalized efficiency as function of mesh size is shown in Fig. \ref{Fig_Mesh_analysis}, where the mesh used for the subsequent calculations have also been indicated.

\section{Heat losses by conduction}
We first consider the isolated case of heat losses only occurring by conduction. This corresponds to the situation where the legs in a TEG are  surrounded by a thermally insulating material with a given thermal conductivity, $\kappa$, and an infinite electrical resistivity. We assume that no heat is lost through the outer sides of the module. As the temperature profiles in the n- and p-legs are almost identical in the no heat loss case, and as the legs have the same size, the efficiency is the same regardless of both the number of legs and the absolute size of the module.

\begin{figure}[!t]
  \centering
  \includegraphics[width=1\columnwidth]{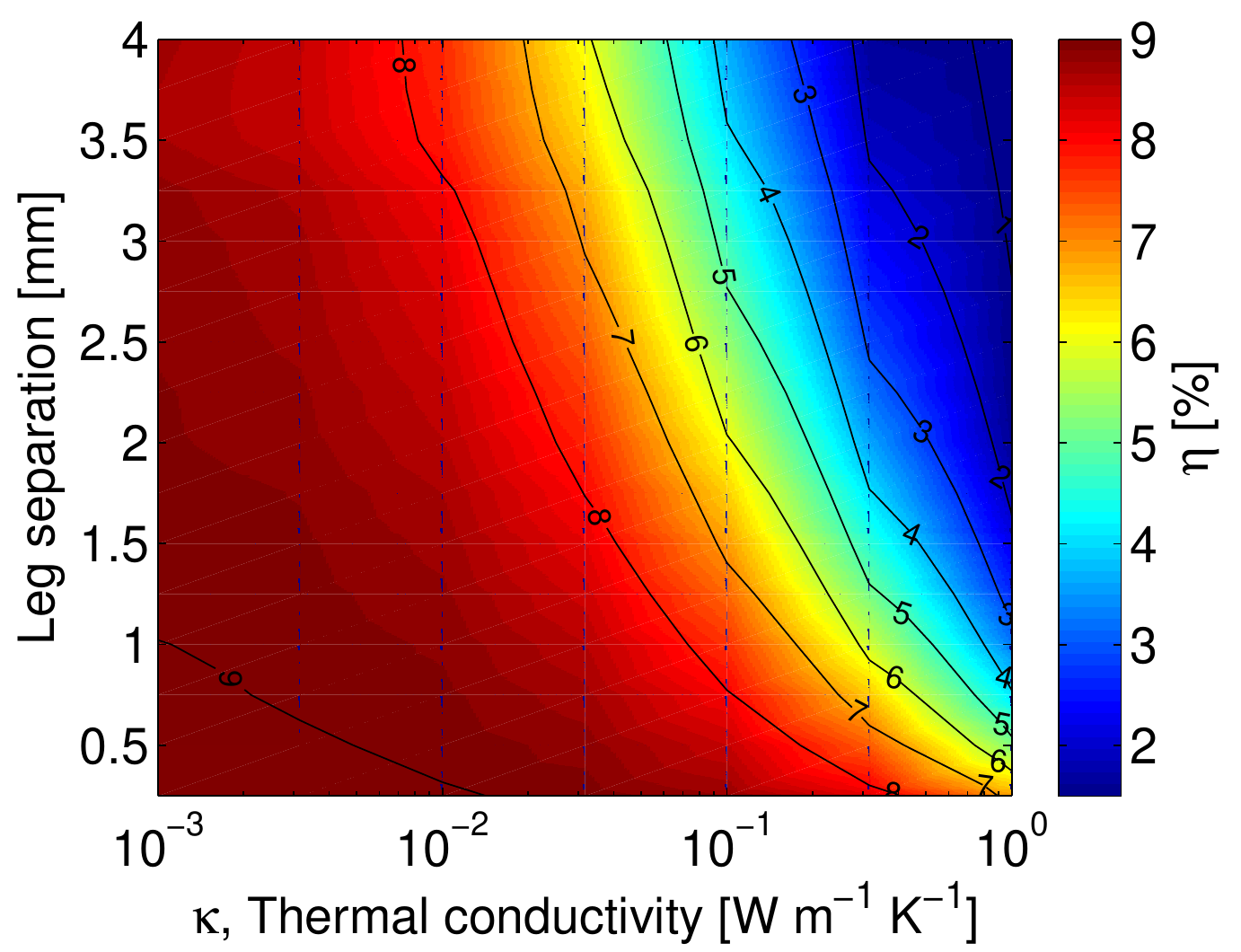}
  \caption{A surface plot of the efficiency, $\eta$, as function of the leg separation distance and the thermal conductivity of the insulator material. The efficiency is independent of the number of legs in the module.}
  \label{Fig_Surf_efficiency_leg_sep_1}
\end{figure}

The parameters varied for this system were leg separation values of 0.25-4 mm in steps of 0.25 mm and $\kappa = 10^{x}$ Wm$^{-1}$K$^{-1}$ from $x$ = $-3$ to $0$ in steps of $0.5$. Shown in Fig. \ref{Fig_Surf_efficiency_leg_sep_1} is a surface plot of the efficiency as function of the leg separation distance and the thermal conductivity of the insulator material. The efficiency decreases for both increasing thermal conductivity and leg separation. This is expected as higher thermal conductivities and leg separations reduce the thermal resistance of the insulator material and thus increases the heat flux that bypasses the legs. Note that the thermal conductivity of the p- and n-leg materials is around 1 Wm$^{-1}$K$^{-1}$ for the materials considered here. For the thermal conductivity of still air, $\kappa\approx 0.02-0.04$ W m$^{-1}$ K$^{-1}$ for the temperature range considered here \citep{Lemmon_2000}, the efficiency is $8-9\%$ for small leg separations. Fig. \ref{Fig_Surf_efficiency_leg_sep_1}  can be used to determine the optimal insulator material if e.g. there is a significant price difference between two insulator material, and the cost per efficiency loss ratio is known.

\section{Heat losses by radiation}
\begin{figure}[!t]
  \centering
  \includegraphics[width=1\columnwidth]{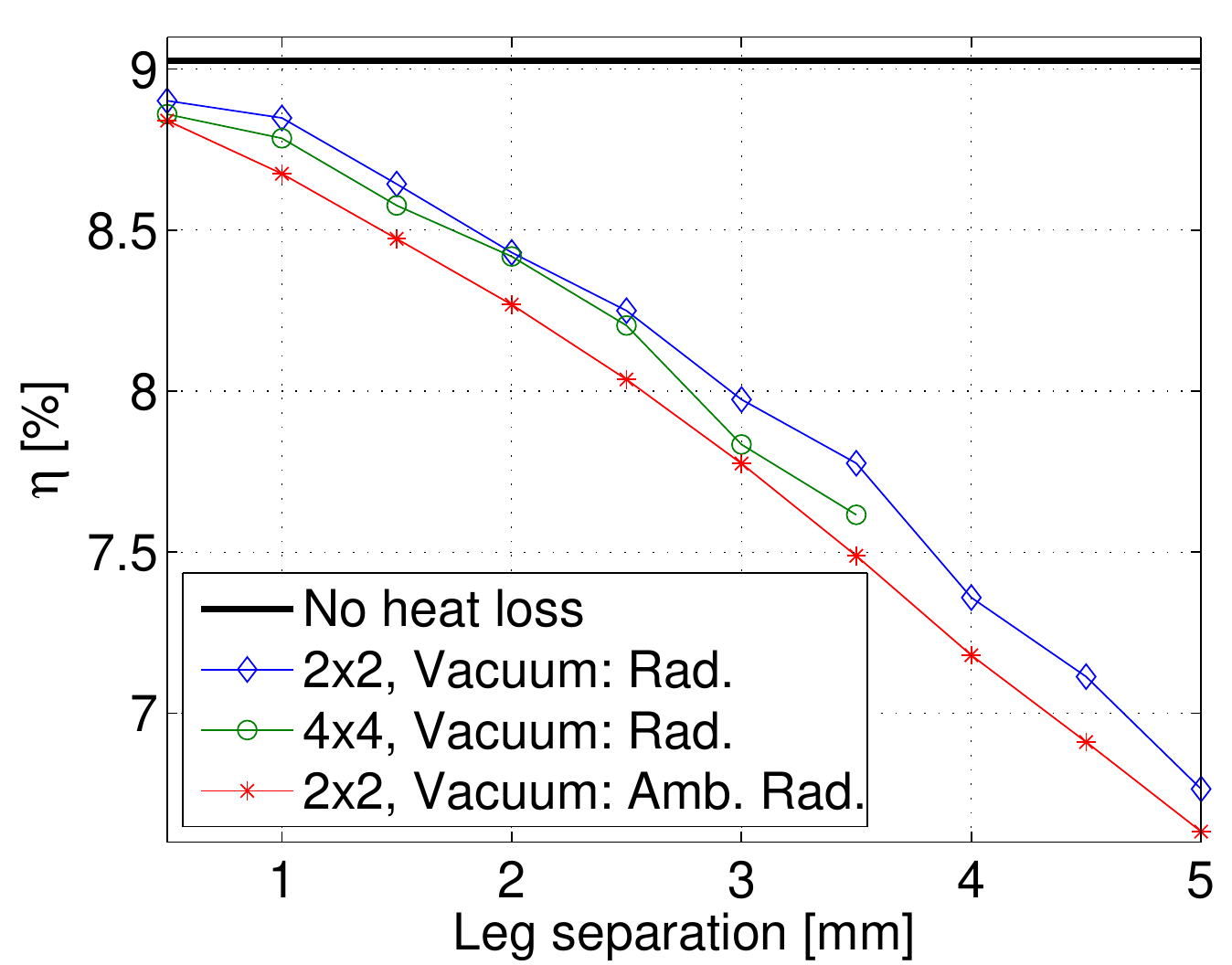}
  \caption{The efficiency, $\eta$, a function of the leg separation distance for a 2x2 and 4x4 module including surface to surface radiation and for a 2x2 module including only surface to ambient radiation. Due to computational resources, only leg separations up to 3.5 mm are considered for the 4x4 module.}
  \label{Fig_Compare_heat_losses_rad}
\end{figure}

We next consider the case of heat loss by radiation only. In this configuration the sides of the module are modelled as diffuse mirrors, and the module is assumed completely evacuated. We consider modules of sizes 2x2 and 4x4, to evaluate the influence of increasing leg number. Also, in order to compare the importance of including surface to surface radiation, the efficiency of the 2x2 and 4x4 modules are compared to that of a 2x2 module, where all surfaces are assumed to radiate only to the ambient. The efficiency as function of leg separation distance is shown in Fig. \ref{Fig_Compare_heat_losses_rad}. As can be seen from the figure, the efficiency of the 2x2 and 4x4 modules are comparable, with the largest difference being less than 0.2\% efficiency. This means that the heat loss computed for the 2x2 model can be considered a good approximation of an $n\times{}n$ system. The difference between including surface to surface and only surface to ambient radiation is also seen to be small, on the order of 0.35\% efficiency. This means that the computationally much faster surface to ambient method can be used to estimate the radiative heat loss, at least for the temperatures considered here.

\section{Heat losses by convection}
Heat loss by convection have been considered previously in the literature for a TEG, by specifying a heat transfer coefficient and not by modelling the actual fluid flow \citep{Chen_2011,Ziolkowski_2010,Reddy_2013,Bauknecht_2013,Yang_2011,Saber_2002}. This approach is a simplification of the true physical problem, in which the heat transfer coefficient must depend on the position on the surface of the leg(s). Here we consider the losses due to natural convection inside a closed and sealed module, filled with air. The module is assumed filled with atmospheric air at atmospheric pressure, with temperature dependent properties, thus the flow is assumed to be compressible. The air is considered transparent to radiation. Transport of heat can thus occur both by radiation between surfaces (of different temperature), conduction through the air and convection. The flow of air is assumed to be laminar.

The hot and cold surfaces of the TEG will establish a gradient in density, which can lead to convective cells. The balance of forces is between the gravitational force on the gradient in density as opposed to the viscous damping force in the fluid. This balance is expressed by the non-dimensional Rayleigh number, $\n{Ra}_\n{L}$, defined as
\begin{eqnarray}
\n{Ra}_\n{L}=\frac{g\beta}{\nu{}\alpha{}}(T_\n{b}-T_\n{t})L^3
\end{eqnarray}
where $g$ is the acceleration due to gravity, $\nu$ is the kinematic viscosity, $\alpha$ is the thermal diffusivity, $\beta$ is the thermal expansion coefficient, $T_\n{b}$ and $T_\n{t}$ is the temperature of the bottom and top plates, respectively, and $L$ is the height of the container.

The critical Rayleigh number depends on the boundary conditions of the system. For the rigid-rigid boundary system of the TEG module, the critical Rayleigh number depends on whether there is vertical velocity symmetry with respect to the mid plane or vertical velocity antisymmetry. The former mode has one row of cells along the vertical while the latter has two row of cells. The critical Rayleigh numbers are 1707 and 17610 for the two cases, respectively.

For the TEG module, placed with the cold side on top and the hot side on the bottom, the Rayleigh number is on the order of 10-30 for the temperature range and size of module considered here. This means that no convection cells will develop inside the TEG. However, a convective flow can be established inside the TEG if the TEG is placed such that the hot and cold sides are vertical. In this configuration the air will be heated as it raises along the hot vertical surface and cooled as is falls along the cold vertical surface, allowing a convective flow to be established. This configuration means that the individual air currents are in maximal contact with the heating and cooling surfaces, resulting in the largest convection possible. The efficiency of this configuration will be examined below.

\section{Comparing the heat loss mechanisms}
We now turn to studying the efficiency as function of leg separation for a combination of the different heat loss mechanisms described above. We consider the efficiency as function of the leg separation for the following cases
\flushleft
\begin{itemize}
\item Radiation losses for evacuated module.
\item Conduction losses for a module filled with opaque still air ($\kappa = 0.035$ Wm$^{-1}$K$^{-1}$) or still argon ($\kappa = 0.016$ Wm$^{-1}$K$^{-1}$).
\item Conduction and radiative losses for a module filled with still transparent air.
\item Convection, conductive and radiative losses for a vertical module filled with transparent air.
\end{itemize}

\begin{figure}[!t]
  \centering
  \includegraphics[width=1\columnwidth]{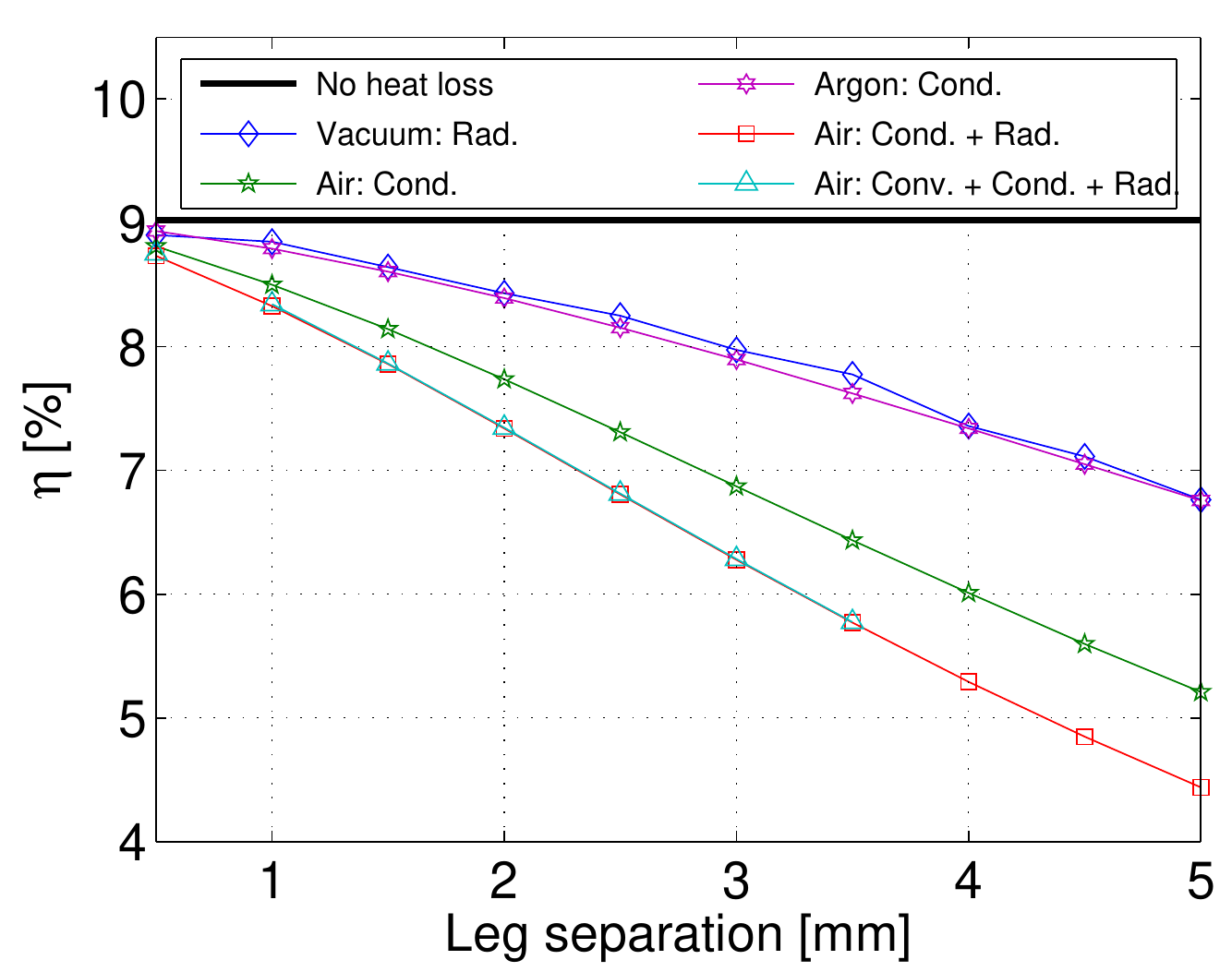}
  \caption{The efficiency, $\eta$, as function of leg separation for the different heat loss cases. Due to computational resources, only leg separations up to 3.5 mm are considered for the convective heat loss case.}
  \label{Fig_Compare_heat_losses}
\end{figure}

For all these cases the efficiency as a function of leg separation is shown in Fig. \ref{Fig_Compare_heat_losses} for the optimal load resistance. The largest heat loss occurs for the combined case of conduction and radiation. Convection is seen to result in no additional heat loss, even though a nice circular flow can be established inside the TEG, as shown in Fig. \ref{Fig_losses_ill_combined_single}. The flow velocity is a few mm/s, which, for a module of this size, results in negligible convective heat transfer. The convection results were calculated for a 2x2 module, but convective losses can be expected to be equally negligible for modules with a larger number of legs. Interestingly, the efficiency as function of leg separation is almost the same for a module filled with opaque argon and an evacuated module with only radiation losses. The conclusion is that an efficient module can be designed either with a filled solid insulator or by evacuating the module completely. At higher temperature, radiation losses will increase strongly, as these are proportional to $T^4$, which means that it is critical to include these in TEG module modelling.

\begin{figure}[!t]
  \centering
  \includegraphics[width=0.95\columnwidth]{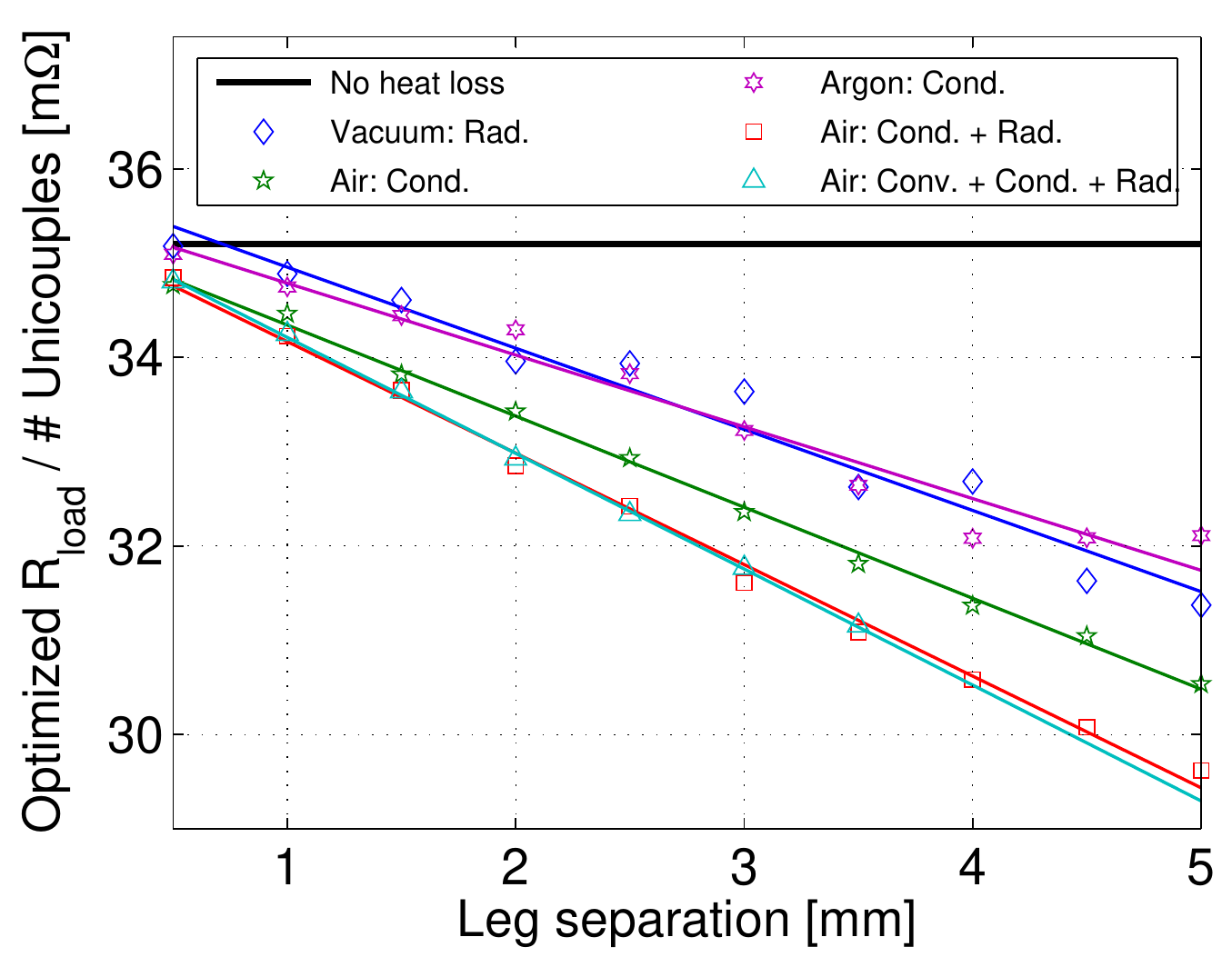}
  \caption{The optimized load resistance per unicouple as function of leg separation for the same systems as considered in Fig. \ref{Fig_Compare_heat_losses}. The lines are straight lines fitted to the data.}
  \label{Fig_Compare_heat_losses_res}
\end{figure}

It is also of interest to consider the optimal load resistance for the different heat loss cases. As previously mentioned, the optimal resistance has been determined for every heat loss mechanism and for all the parameters considered in this study. The resistance varies with leg separation, and thereby heat loss, in a systematic way, decreasing as the leg separation increases. This can be seen in Fig. \ref{Fig_Compare_heat_losses_res}. It is numerically difficult to exactly determine the optimal load resistance, as the efficiency is a shallow function of the load resistance near the optimum value. This results in a small variance in the optimal determined resistances shown in the figure. However, it seems clear that the decrease in load resistance is linear with the leg separation. The reason for this decrease is that as the heat losses increase, it becomes  beneficial to operate the TEG at maximum power, rather than at maximum efficiency. Therefore the resistance will go towards the resistance for optimal power, which for the geometry considered above is at 25.6 m$\Omega$ per unicouple.

\section{Heat loss for different leg dimensions}
\begin{figure}[!t]
  \centering
  \includegraphics[width=0.95\columnwidth]{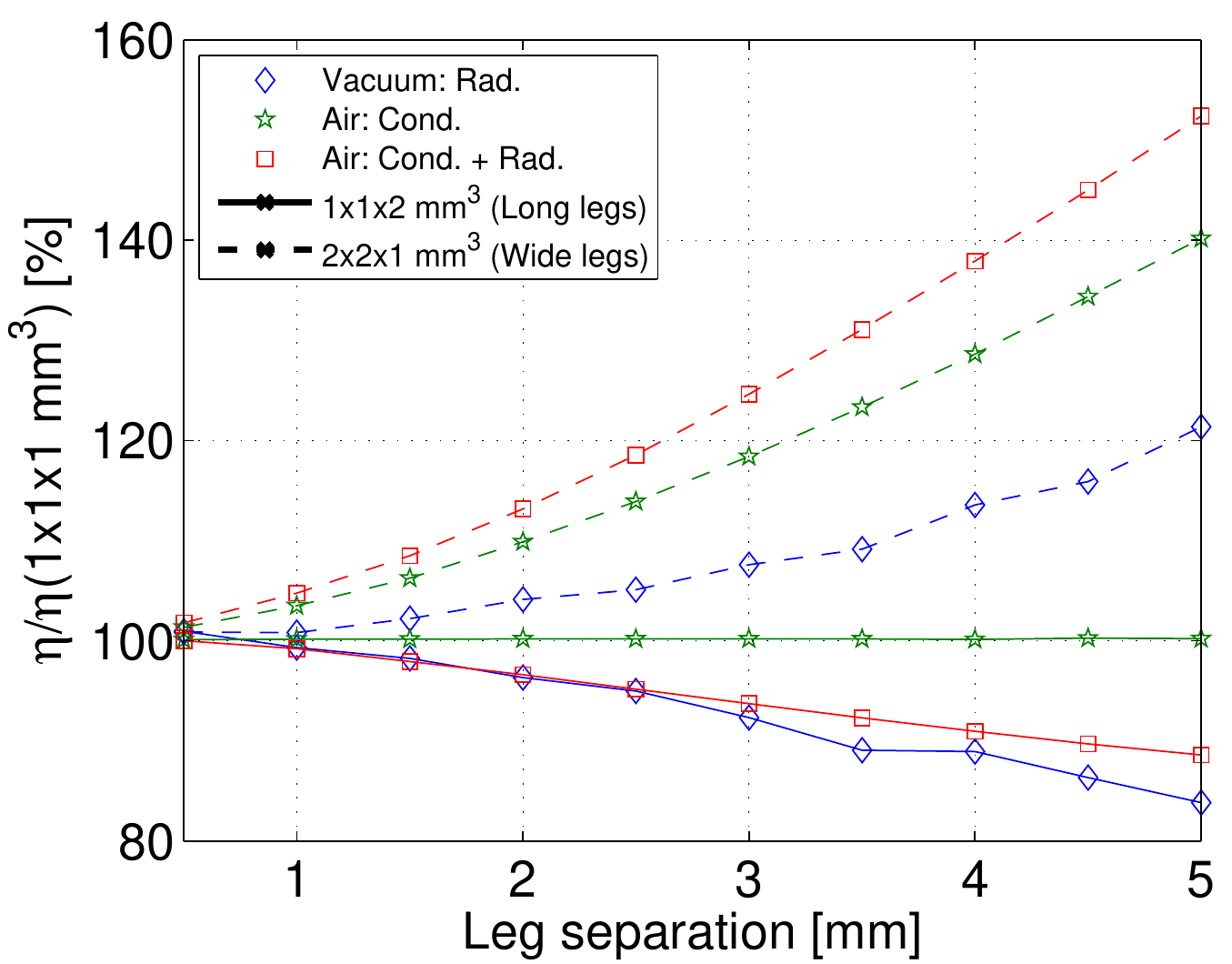}
  \caption{The efficiency, $\eta$, as function of leg separation for the different heat loss cases and for different leg dimensions normalized with the $1\times{}1\times{}1$ mm$^3$ system. The argon case considered in Fig. \ref{Fig_Compare_heat_losses} is not shown, but results are in trend similar to the case of air.}
  \label{Fig_Compare_heat_losses_long_wide}
\end{figure}

To determine the heat loss as function of the dimensions of the legs, the efficiency has been computed for modules with wide legs with leg cross-sectional areas, $(x_\n{leg},y_\n{leg})$, of 2 mm $\times{}$ 2 mm and a leg length, $z_\n{leg}$, of 1 mm, and for long legs with leg cross-sectional area of 1 mm $\times{}$ 1 mm and a leg length of 2 mm. As previously, the size of the n- and p-leg is equal. These are wider and longer legs, respectively, than the 1 mm $\times{}$ 1 mm $\times{}$ 1 mm legs considered above. The normalized efficiency as function of leg separation distance is shown in Fig. \ref{Fig_Compare_heat_losses_long_wide} for the different heat loss cases. When only conduction is considered, increasing the leg length and with a fixed cross-sectional area the efficiency remains constant. This is expected due to the adiabatic boundary conditions, as well as the almost linear temperature profile along the legs, which means that heat loss from the leg to the air is negligible. In the radiation case, longer legs lead to larger heat loss, as the legs have a larger surface from which to radiate heat to the cold plate.  The wide legs are seen to substantially increase the efficiency for both the radiative and conductive heat loss cases. This is due to a smaller surface to cross-sectional area ratio, which leads to lower heat loss compared to the amount of heat flowing through the leg. The use of wider legs is seen to be increasingly beneficial at larger leg separations, due to the increased cross-sectional area of the legs compared to that of the top plate.

The decrease in efficiency for longer or larger legs can be understood by noting that heat losses have the least effect on the efficiency if the majority of the heat flows through the legs and only a minor fraction bypasses the active material via the heat loss mechanisms. Scaling the absolute leg size with a factor $x$, increases the conduction flow through the legs linearly as the cross-sectional area increases quadratically, but the length increases linearly. However, the leg surface area – where the heat losses occur – scales quadratically. Thus, when up-scaling the legs, heat losses becomes more dominant. Instead, shorter legs and large cross-sectional areas are optimal in reducing heat losses.

\begin{figure*}
\begin{center}
\subfigure{\includegraphics[width=0.49\textwidth]{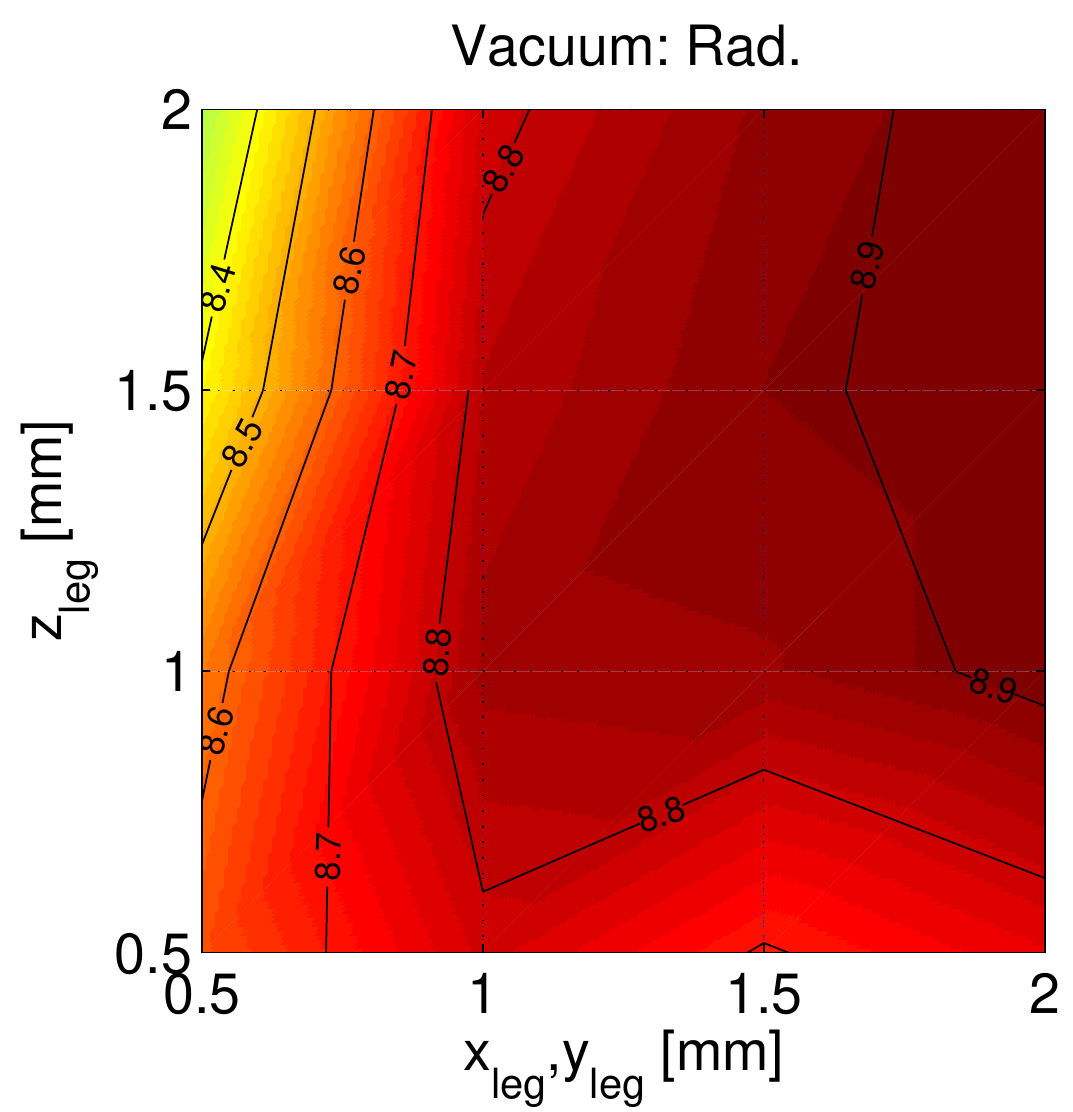}}\hspace{0.1cm}
\subfigure{\includegraphics[width=0.49\textwidth]{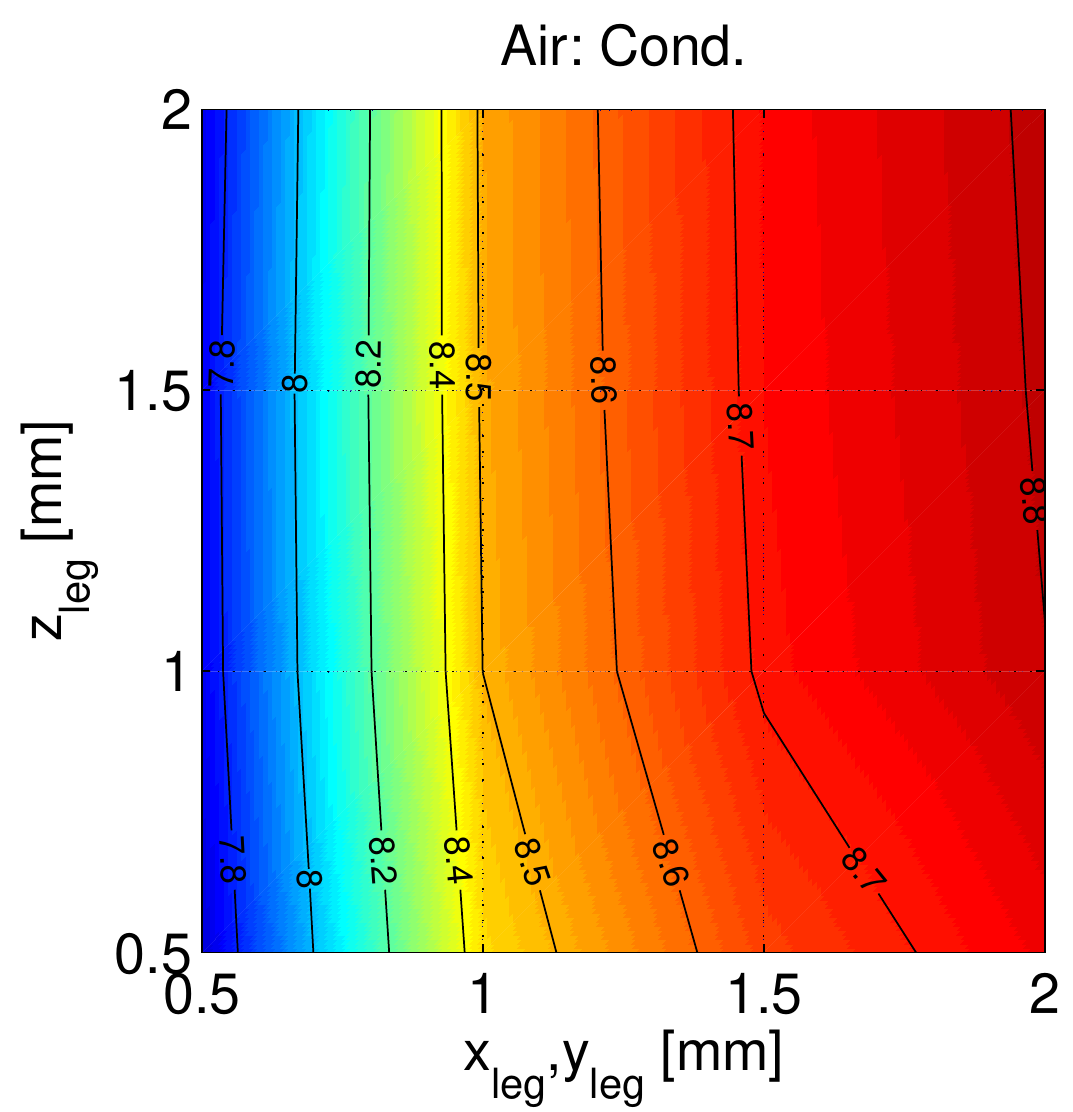}}
\subfigure{\includegraphics[width=0.49\textwidth]{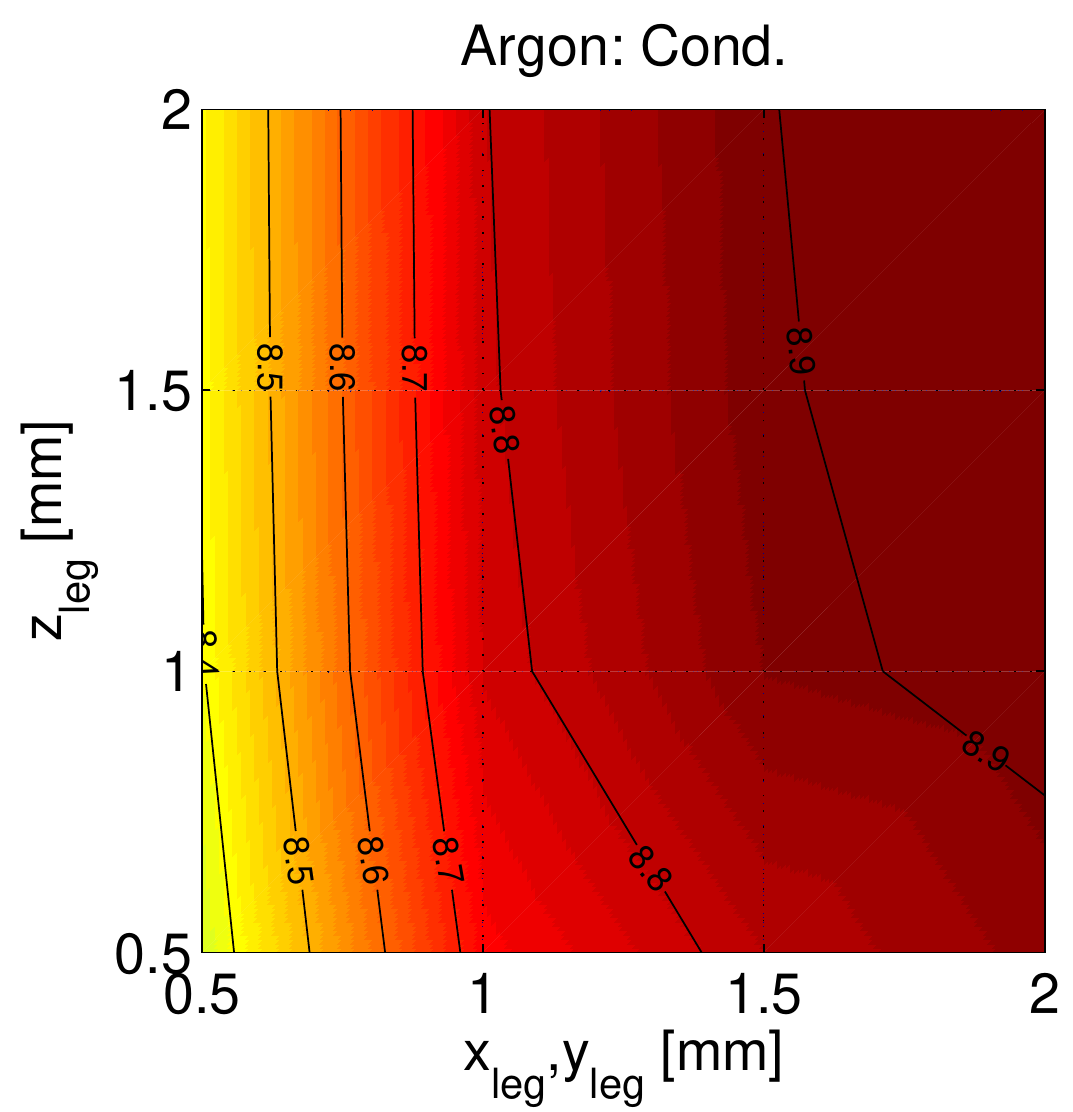}}\hspace{0.1cm}
\subfigure{\includegraphics[width=0.49\textwidth]{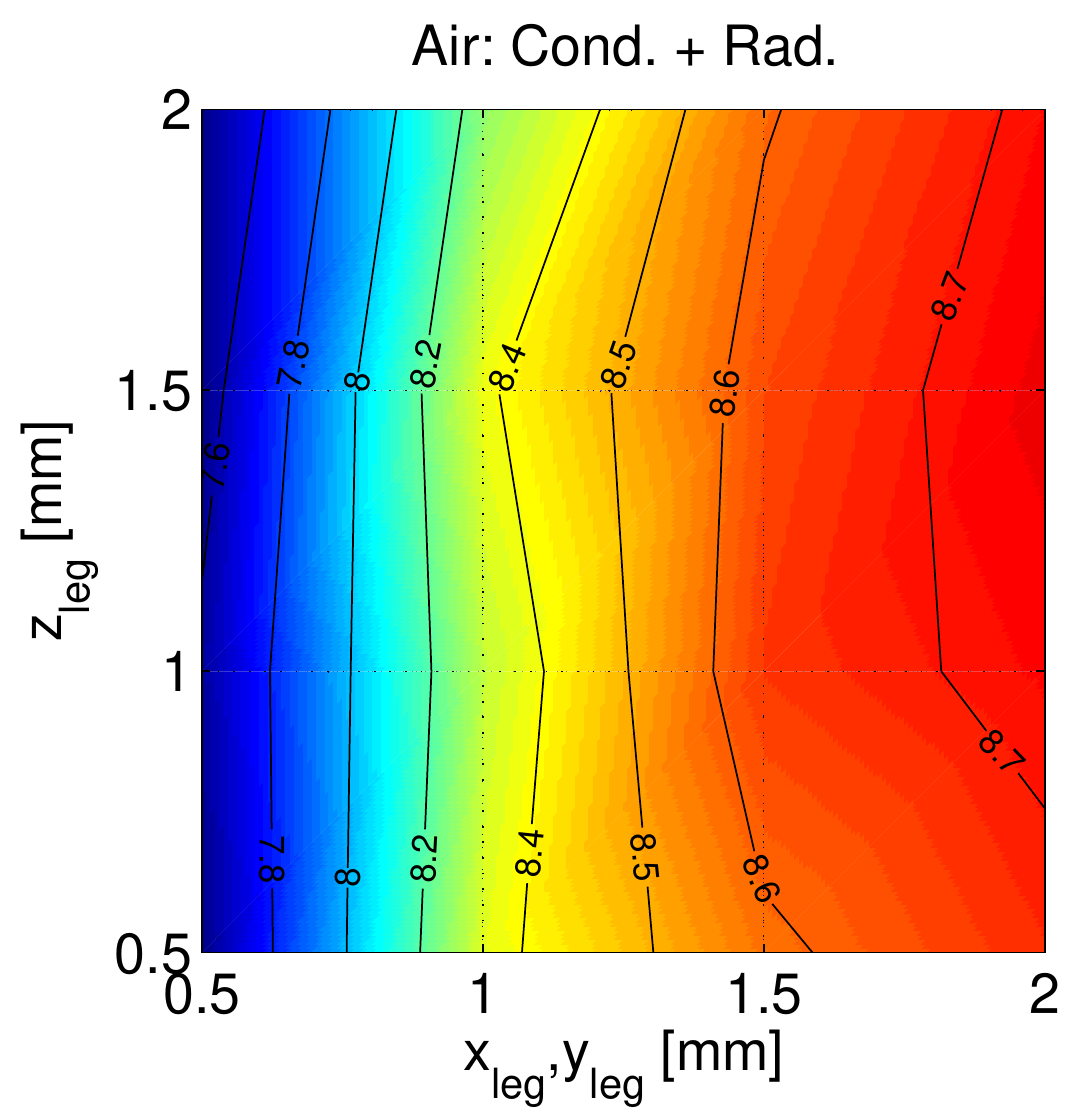}}
\caption{The efficiency, $\eta$, as function of leg dimensions in the $x_\n{leg}=y_\n{leg}$ and $z_\n{leg}$ dimensions, respectively, for a leg separation value of 1 mm, i.e. regardless of the dimensions of the legs there is always 1 mm between the legs. The colour scale is the same for the different heat loss mechanisms. The results are interpolated based on values calculated from 0.5 mm to 2 mm in steps of 0.5 mm. The radiation in vacuum case has a slight numerical variation in the value for the efficiency.}
  \label{Fig_Dim_heat_loss}
\end{center}
\end{figure*}

The heat loss as function of the leg cross-sectional area, $x_\n{leg}=y_\n{leg}$, and leg length, $z_\n{leg}$, has also been considered for a fixed leg separation of 1 mm. Fig. \ref{Fig_Dim_heat_loss} show the efficiencies of the module for different legs dimensions and heat losses based on calculated leg values of 0.5-2 mm in steps of 0.5 mm. As seen in Fig. \ref{Fig_Dim_heat_loss}, increasing the legs cross-sectional area, $(x_\n{leg},y_\n{leg})$ increases the efficiency of the module while changing the length of the legs only decreases the efficiency in the case of radiative heat loss. This is as observed in Fig. \ref{Fig_Compare_heat_losses_long_wide}. Note that, as previously mentioned, had the legs had different temperature profiles due to different material properties, the legs will start to cross-talk thermally at small separations in the conduction case also. The ideal way to construct a TEG module with minimal heat losses is thus to use a good insulating material between the legs, properly seal the module and using small and wide, closely spaced legs.

\section{Conclusion}
A three dimensional numerical finite element model of a thermoelectric generator consisting of p- and n-leg bismuth telluride has been developed. The model was compared with a one dimensional numerical model for the case of no heat losses and the results were shown to be identical. Next, different heat loss mechanisms were investigated for a closed module. Both surface to surface radiative heat transfer, conductive and convective heat transfer were considered, and both separately and simultaneously. First, the influence of surface to surface radiation was shown to be small for the hot side temperature of 523.15 K considered here and using the simpler surface to ambient radiation is sufficient. Second, for conductive heat transfer the efficiency as function of thermal conductivity and unicouple separation was determined. Third, heat losses due to convection inside the module were shown to be negligible for the module size considered here. Comparing the heat loss mechanisms, it was shown that for an insulator with properties similar to argon, the efficiency is comparable to radiative heat losses. The efficiency was significantly lower for combined radiative and conduction heat transfer. Investigating the dimensions of the legs showed that the ideal way to construct a TEG module with minimal heat losses is to either use a good insulating material between the legs or evacuate the module completely, and use small and wide legs closely spaced.

\section*{Acknowledgements}
The authors would like to thank the Programme Commission for Energy and Environment (EnMi), The Danish Research and Innovations (Project No. 10-093971) for sponsoring the OTE-POWER research work.

\end{document}